\documentclass[12pt,usenatbib,preprint]{aastex}

\newcommand{\bfig}{\begin{figure}}
\newcommand{\efig}{\end{figure}}
\newcommand{\bfige}{\begin{figure*}}
\newcommand{\efige}{\end{figure*}}
\newcommand{\bea}{\begin{eqnarray}}
\newcommand{\eea}{\end{eqnarray}}
\newcommand{\md}{M$_{\odot}$~}

\newcommand{\chandra}{{\it Chandra}}

\shorttitle{M33 X-ray binary population}
\shortauthors{Grimm et al.}

\begin{document}

\title{The X-ray binary population in M33: I. Source list and
luminosity function}
\author{H.-J. Grimm\altaffilmark{1}, J.~McDowell\altaffilmark{1},
A. Zezas\altaffilmark{1}, D.-W. Kim\altaffilmark{1},
G.~Fabbiano\altaffilmark{1}}
\affil{Harvard-Smithsonian Center for Astrophysics, 60 Garden Street,
Cambridge, MA 02138}

\begin{abstract}
In this paper we present the source list for 3 \chandra~observations
of the Local Group galaxy M33. The observations are centered on the
nucleus and on the star forming region NGC 604. We detect a total of
261 sources in an area of $\sim$ 0.2 square degree down to a flux
limit of $3\times 10^{-16}$ erg s$^{-1}$ cm$^{-2}$ which corresponds to
a luminosity of $\sim 2\times 10^{34}$ erg s$^{-1}$ at a distance of
840 kpc. From the source list we construct the luminosity functions
of sources observed in M33. Taking into account background
contamination the luminosity functions are consistent with those of
other star forming galaxies. In addition, the combination of X-ray
color analysis and the existence of ``blue'' optical counterparts
strongly indicate that the X-ray point source population in M33
consists of young objects. Above $3\times 10^{35}$ erg s$^{-1}$ there
are few X-ray sources in the locus of the X-ray hardness ratio diagram
that is generally populated by LMXBs.
\end{abstract}


\section{Introduction}
\label{sec:intro}

M33 is a late-type spiral galaxy, type Sc II-III, and the
third largest galaxy in the Local Group. In the Local Group it
occupies a unique place since morphologically it is of intermediate
type between the large early-type spiral galaxies (Milky Way and M31)
and the numerous irregular galaxies. In contrast to M31 and the Milky
Way, M33 does not have a stellar bulge, and also does not contain a
super-massive black hole at the center \citep{gebhardt:01}. Other
galaxies of this type cannot be investigated with the same depth even
with \chandra: at a distance of 840 kpc \citep{freedman:91} M33 is the
second nearest spiral galaxy; it spans roughly 73$\times$45 arcminutes
on the sky; the line-of-sight absorption column density is small, $N_H
\sim 6  \times 10^{20}$ cm$^{-2}$\footnote{obtained from the colden
tool (http://cxc.harvard.edu/toolkit/colden.jsp) based on NRAO  data
compiled by \citet{dickey:90}}. M33 is more actively star forming (SFR
$\sim$ 0.3--0.7 \md yr$^{-1}$) than either the Milky Way or
M31 \citep{hippelein:03}, particularly compared to its much smaller
mass. This star formation activity is visible in a number of large HII
regions in the disk of M33.

As a large nearby galaxy, M33 has been observed with a number of X-ray 
satellites. The Einstein observatory detected 17 point sources
\citep{trinchieri:88}, among them a clear eclipsing binary (M33 X-7)
\citep{peres:89}. \citet{markert:83} established the 
existence of an X-ray binary population based on X-ray variability and 
noted the association of the X-ray sources with young stellar populations.
ROSAT observations led to the detection of a total 184 fainter sources 
coincident with Population I markers; a number of these X-ray sources
are identified with SNRs \citep{long:96,haberl:01}. BeppoSAX
spectroscopy showed that the spectra of the brighter X-ray sources in
M33 are consistent with X-ray binaries, and that up to 5 sources are
variable \citep{parmar:01}.

M33's proximity, angular size, and low line-of-sight column
density provide unique advantages for the study of the X-ray
population of this late type spiral galaxy: 1) The same amount of
detail is achievable in M33 as in M31 (1'' = 4.1 pc). 2) The smaller
angular size of M33 allows the observation of a larger fraction of the
galaxy in a single exposure, unlike M31 or LMC/SMC. 3) The low
line-of-sight $N_H$ and the proximity of M33 allow the study of a
luminosity range (few $10^{34} - 10^{38}$ erg s$^{-1}$) unaccessible
in more distant galaxies. Sources in this luminosity range are more
difficult to survey in both the Milky Way and M31 because of their larger
angular sizes requiring very long observing times. Moreover, in the Milky Way
distance uncertainties for individual sources, and partly high $N_H$
make this type of study more uncertain. 4) The higher star formation
rate of M33 results in a different population of X-ray sources
compared to those in Milky Way and M31. 5) Its proximity combined with
the positional accuracy and sensitivity of \chandra~allows a detailed
comparison of the X-ray sources with cataloged sources at other
wavelengths. This is crucial to understand the X-ray source population
in detail and to gain a complete picture of the X-ray binary population.

In this paper we present the source list, luminosities, X-ray
luminosity functions, and identifications of X-ray point sources in
M33. An accompanying paper will investigate the detailed properties
and X-ray spectra of the point source population.

\section{\chandra~data and Analysis}
\label{sec:data}

For this study we used three of the \chandra~observations of
M33 (Table \ref{tab:obs}). A short ($\sim12$ ks) fourth observation,
ObsId 787, was disregarded. This observation was aimed at studying 
the nucleus and suffers from both high background and small field of
view (FOV) ($\sim$54 square arcmin). Due to the angular extent of M33
all active ACIS chips cover areas of M33. These include the standard
ACIS-S configuration for ObsId 786, and the standard ACIS-I
configuration for ObsIds 1730 and 2023. There is considerable overlap
between the different observations. However, because of the decreasing
resolution/sensitivity  with increasing off-axis angle only the inner
parts of M33 ($\sim$ 8--10 arc minutes) have a significant number of
detected sources in two observations, 786 and 1730.

\begin{table}[h]
\begin{center}
\caption{List of ACIS observations of M33.}
\begin{tabular}{|l|c|r|c|c|c|}
\hline
ObsId & Date & Aim point & Duration [ks] & Chips & Datamode\\
\hline
786        & 2000-08-30 &  Nucleus & 45 & ACIS-I 2--3, ACIS-S 1--4 & FAINT \\
1730       & 2000-07-12 &  Nucleus & 45 & ACIS-I 0--3, ACIS-S 2--3 & VFAINT \\
2023       & 2001-07-06 &  NGC 604 & 90 & ACIS-I 0--3, ACIS-S 2--3 & FAINT \\
\hline
\end{tabular}
\label{tab:obs}
\end{center}
\end{table}

We used CIAO 3.0.1 and 3.0.2 and CALDB 2.24 and 2.25 to analyze the
data. There were no changes to the ACIS data
preparation software or to the source detection software or to
relevant files in the CALDB between these versions.  We do not correct
for enhanced absorption due to the hydrocarbon accumulation on the
chips. The observations do not suffer from strong background flares,
but short times of higher background (3$\sigma$ above the mean) have
been excluded. The effective exposure times are $\sim$45 ks, $\sim$43
ks, and $\sim$89 ks for observations 786, 1730, and 2023,
respectively. We correct for time-dependent variations of the gain
with the {\em corr\_tgain} program developed by Alexey
Vikhlinin\footnote{http://asc.harvard.edu/cont-soft/software/corr\_tgain.1.0.html}.

Source detection was performed with {\em wavdetect} on scales of 1, 2,
4, 8, 10, 12, and 16 pixels. The full energy range, 0.3--8.0 keV,
was also subdivided in a soft, 0.3--2 keV, and hard band,
2--8 keV. Source detection was done in all the three bands. The
chance detection threshold was set to $10^{-6}$, equivalent to 1
spurious detection per million pixels. Due to strong
pile-up, these scales fail to detect the nucleus of M33
unambiguously. Moreover, the HII regions NGC 604 and IC 131, which are 
more extended than the scales used, were detected by visual inspection 
and interactive analysis. The nucleus of M33 is strongly piled up and
was therefore excluded from the analysis. For details on the 
nuclear source refer, e.g. to \citet{dubus:04} and
\citet{laparola:04}. The source regions correspond to the 3$\sigma$
ellipses output from {\em wavdetect}. All source regions
were inspected by eye and, if necessary, adjusted for overlapping with 
nearby regions and to cover the whole extent of a source. Due to the
partly crowded fields, particularly near the aim point, we chose
background regions on each chip in a large source-free area to ensure
a sufficient number of background counts. This background region was
used for all sources on the corresponding chip. Since M33 has no
significant intrinsic diffuse emission due to hot gas, with the
exception of the inner 2 arcminutes around the nucleus, the background
does not vary significantly over our FOV. Therefore our approach does
not introduce significant errors in the intensity of the sources. Note 
that for ObsId 786 most of the sources are on the back-illuminated (BI)
ACIS-S3 chip, whereas the majority of sources in the other
observations is on the ACIS-I chips. The source counts were computed
according to: 
\begin{equation}
  C_{source} = C - \frac{A_{source}}{A_{bkg}} \cdot C_{bkg} \cdot \frac{E_{source}}{E_{bkg}}
\end{equation}
where $C_{source}$ is the source counts, $C$ is the observed counts,
$A_x$ are source and background region areas of the extraction
apertures, $C_{bkg}$ is the background counts, and $E_{source,bkg}$
are the exposures for the source and background areas. For sources that
are not detected in one of the energy bands we compute upper limits on
the number of source counts following the approach of
\citet{kraft:91}. A few of these sources show non-zero lower limits
which indicates the presence of a source. In these cases we used the
most probable value for the source counts obtained with the method of
\citet{kraft:91}. These sources are marked in Table \ref{tab:sources}.

In order to obtain proper positional uncertainties for our sources,
which are important for identifications with objects at other
wavelengths, we we use the formula of \citet{kim:04a}. This formula
assumes that variations in the absolute astrometric accuracy
from observation to observation are small and therefore that
the uncertainties depend only on count rate and off-axis angle.
The positional uncertainty is given as a function of off-axis angle
in the high and low count regimes. We use 100 counts to discriminate
between low and high count sources. The density of point sources is
small enough that sources are identified unambiguously, with the
possible exception of the nucleus and the HII region NGC 604. These
two cases, however, are problematic only in terms of the background
they provide for other sources. In the cases were a source is observed
in two or three observations, we take as the positional uncertainty the
smallest value of the available observations.

To derive X-ray colors we subdivided the observations into three
energy bands, soft (0.3--1.0 keV), medium (1.0--2.1 keV), and hard
(2.1--8.0 keV) and measured the number of counts within the source
aperture in each band. For count extraction we use the source regions
in the total band from 0.3--8.0 keV. Since very faint sources,
i.e. sources with less  than 20 counts in the total band, are not used
in the color analysis, this choice does not introduce a significant
error in the extracted counts. Of the 288 sources detected in the
whole band in all observations, 99 sourcces have less than 20
counts. Most of the sources are too faint to be investigated for
spectral variability. Spectral variability of brighter sources for
which spectra can be extracted will be discussed in a separate paper.

\section{Point source detections}

The number of sources detected with {\em wavdetect} in the different
energy bands is given in Table \ref{tab:number}. Note that because of
the overlapping fields of view some of the sources appear in more than
one observation. Also some sources are detected only in one of the
three energy bands. The combined number of individual sources from all 
observations is 261. Among these sources 25 are detected only in the
soft (0.3--2.0 keV) band and 16 only in the hard (2.0--8.0 keV)
band. These hard sources have either intrinsically hard spectra or are 
attenuated by the ISM within M33, since the Galactic absorption
towards M33 is relatively small. Fig.\ref{fig:m33} shows a DSS image
of M33 overlayed with the source regions and the detector outline for
the three observations.
\bfig
  \resizebox{0.5\hsize}{!}{\includegraphics{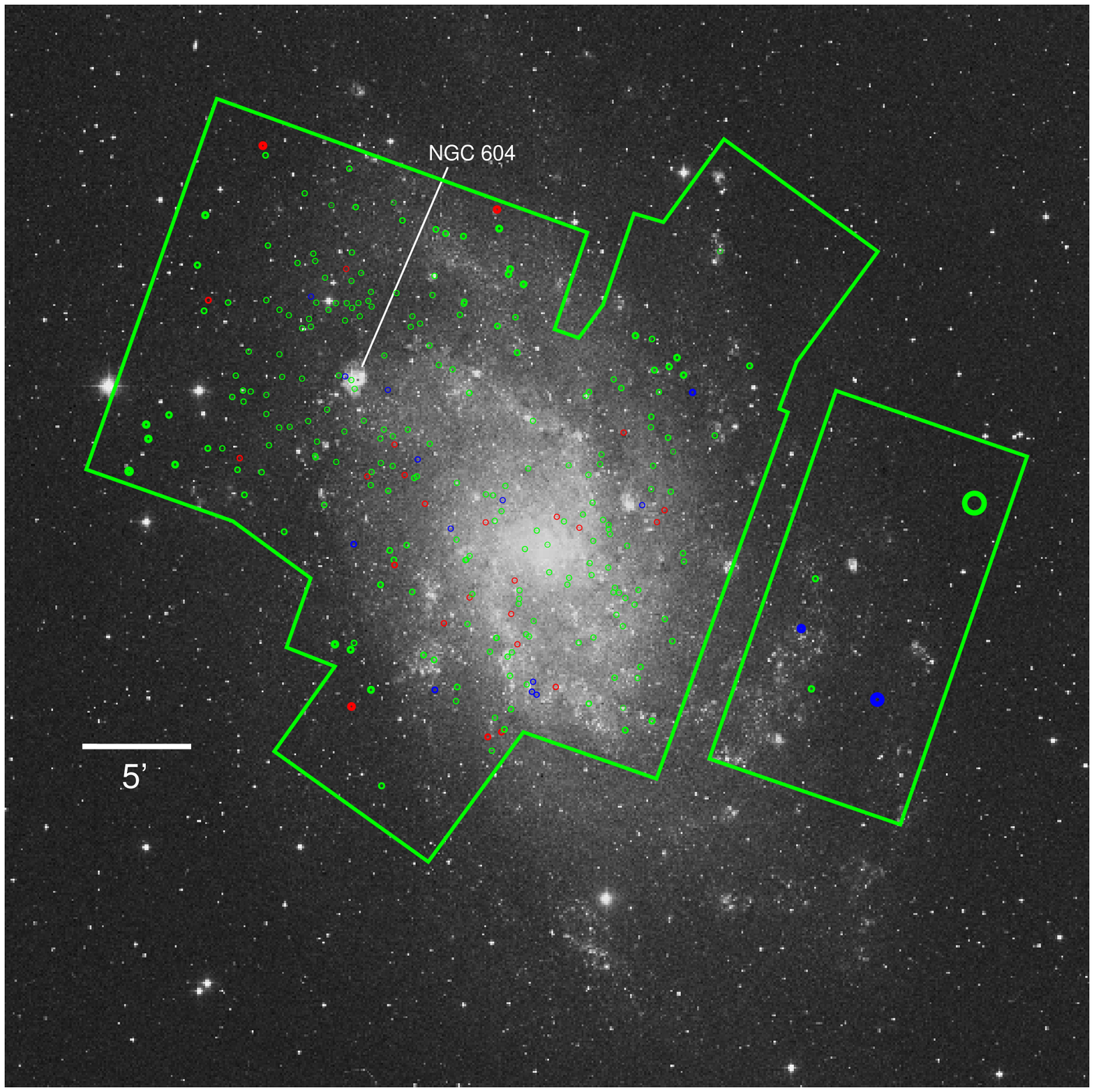}}
  \caption{DSS image of M33 overlayed with source regions and the envelope
    of the \chandra~field of view. North is up, and east is left.}
  \label{fig:m33}
\efig

The source list, including position, counts, and fluxes is
presented in Table \ref{tab:sources} in the Appendix: Column (1) gives
the \chandra~source name, column (2) the sky position, column (3) the
positional uncertainty, columns (4), (5), and (6) the number of counts
in the total, soft, and hard band for the three observations, columns
(7), (8), and (9) the flux in the total, soft, and hard band for the
three observations with 1$\sigma$ errors. The errors given on the
counts are computed using the Gehrels approximation, $\sigma =
1+\sqrt{N+0.75}$ \citep{gehrels:86}. 99\% upper limits on the 
counts were calculated in one of two ways: 1) Because observations 
1730 and 786 have the same aim point the extent of sources is
comparable between these observation; thus if a source was
detected in observations 1730 or 786 in another energy band, this
source region was used to compute the upper limits for observations
1730 and/or 786. 2) If the source was not detected in either of these
observations, upper limits were calculated from counts in a 2
arc second region around the source position.

\begin{table}[h]
\caption{Number of sources detected in different energy bands and
observations. Note that due to overlapping observations some of the
sources appear in more than one observation. Their number for each
band is given in parentheses.}

\begin{tabular}{lcccc}
\hline
ObsId & \multicolumn{3}{c|}{Energy band [keV]} & Total No. of sources\\
\hline
      &  0.3 -- 2.0 & 2.0 -- 8.0 & 0.3 -- 8.0 & \\
\hline
786   &  79 (53) & 47 (32) &  96 (60) & 121\\
1730  &  68 (37) & 48 (30) &  84 (47) & 98\\
2023  &  99 (30) & 68 (21) & 112 (33) & 129\\
\hline
\end{tabular}
\label{tab:number}
\end{table}

The observation time of ObsId 2023 is twice as long as in the other
two observations. This explains the larger number of sources compared
to observation 1730 which uses the same instrumental configuration.
The similar number of source detections in 786 and 2023 is both due to 
the use in 786 of the two back-illuminated (BI) chips
S3 and S5 that have a higher sensitivity at low energies than the
front-illuminated (FI) chips, and to the aim at the center of M33
where the source density might be higher. Because of the
different exposure times, upper limits for non-detections in the 2023
observation are smaller than in the other observations. Moreover, the
$\sim$100\% completeness of this observation reaches down to 
$\sim 2-3\times 10^{-15}$ erg s$^{-1}$ cm$^{-2}$ compared to $\sim
6-7\times 10^{-15}$ erg s$^{-1}$ cm$^{-2}$ for both other
observations. The completeness estimate is derived from simulations
for correcting the luminosity function according to
\citet{kim:03}. The correction is based on simulations of the
detection probability of point sources with a range of flux values for
various off-axis angles from the aimpoint. The simulations are
performed with the MARX
simulator\footnote{http://space.mit.edu/ASC/MARX/}. The effects taken
into account in the simulations are the decreasing sensitivity with
increasing off-axis angle, diffuse background (not important here),
uneven sky coverage of the S3 chip and telescope vignetting as
described in the exposure maps.

For sources with more than 80 counts we extracted radial profiles
of the source counts and fitted them with a Gaussian plus
background. We compared the full-width half maximum (FWHM) of the
Gaussian with a Gaussian fit of the point-spread function. For the 59
sources with sufficient counts we find no evidence of extended
emission.

Given the large area covered by the observations, we expect to have
serious contamination by background AGN, in particular at fainter
luminosities. Based on data from the Chandra Deep Field--North (CDF-N)
\citep{alexander:03} we estimate that of order 50\% of the sources
detected with luminosities in the range $10^{34} \le L_X \le 5 \times
10^{35}$ erg s$^{-1}$ are background AGN. \citet{pietsch:04} arrive
at a comparable fraction for the XMM field, which is roughly four
times larger than the area covered by \chandra. The XMM-Newton
observation is on average less deep than the
\chandra~observations. Cosmic variance may change the number of
background AGN slightly but based on data from the CHAMP survey cosmic
variance does not appear to be a strong effect \citep{kim:04b}. A more
detailed estimate of the actual number of background objects requires
a more detailed knowledge of the source properties. Above $L_X \ge 5
\times 10^{35}$ contamination of the luminosity function is small
because the AGN Log(N)--Log(S) is much steeper than the luminosity
function of galactic X-ray sources.

\section{Point source colors}

Hardness--ratio diagrams are a straightforward way to classify sources in
general, especially for sources that do not have sufficient number of counts
to allow spectral fitting \citep{kim:92,prestwich:03}.

From the counts extracted from the three different bands mentioned in
Sec. \ref{sec:data} we construct two colors, a soft color and a hard
color. The soft color is defined as
\begin{equation}
  HR1 = \frac{M-S}{S+M+H}
\end{equation}
and the hard color as
\begin{equation}
  HR2 = \frac{H-M}{S+M+H},
\end{equation}
where $S$, $M$, and $H$ are the counts in a soft (0.3--1.0 keV),
medium (1.0--2.1 keV), and hard (2.1--8.0 keV) energy band,
respectively. The energy bands were selected to achieve an optimal
separation between soft/thermal and hard/power law components
(Prestwich, private comm.). The results for the different observations
are shown in Fig.\ref{fig:color1}. Following \citet{prestwich:03} we
consider here only sources with more than 20 detected counts in the
full band to somewhat reduce statistical errors on the colors. The
errors on the counts are computed with Gehrels approximation
\citep{gehrels:86}; errors on the colors are computed using standard
error propagation. Note that for small number of counts the errors must
not be regarded as Gaussian 68\% limits. We disregard sources that
have upper limits in any of the bands. Typical net counts for the
sources are in the range from 30--50 counts. The detected number of
counts is dependent on the reponse of the CCD. This  difference is
expected to be largest between FI and BI chips. Therefore we plot the
hardness ratios in separate panels. The model tracks in
Fig.\ref{fig:color1} were computed separately for the BI (ObsId 786)
and FI chips (OdsIds 1730 \& 2023). The strong similarity
of the tracks indicates that differences in the CCD response
between our observations do not have a strong effect on the colors, at
least for sources with more than $\sim$20 counts.
\bfige
\begin{center}
  \resizebox{0.3\hsize}{!}{\includegraphics{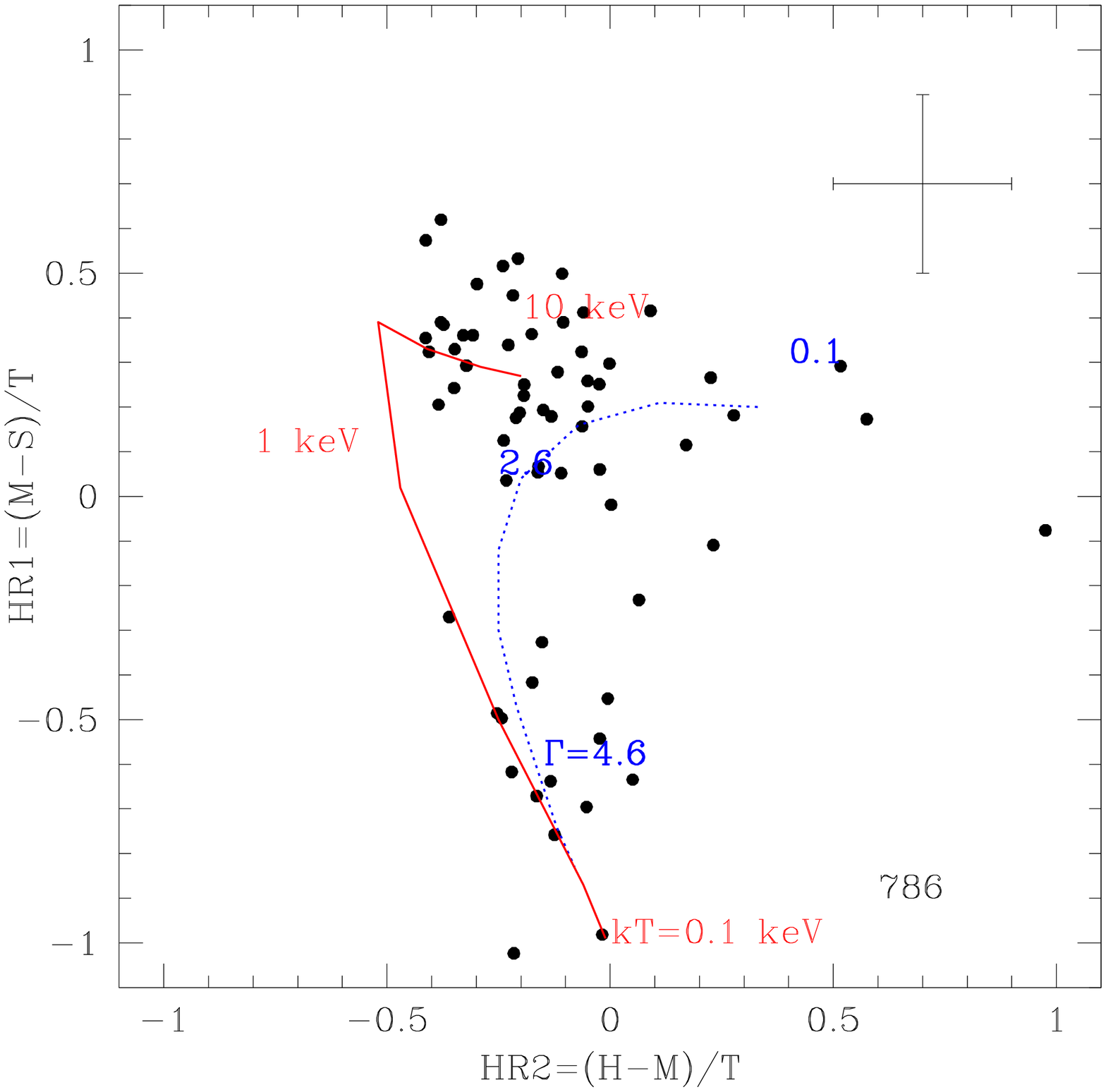}}
  \resizebox{0.3\hsize}{!}{\includegraphics{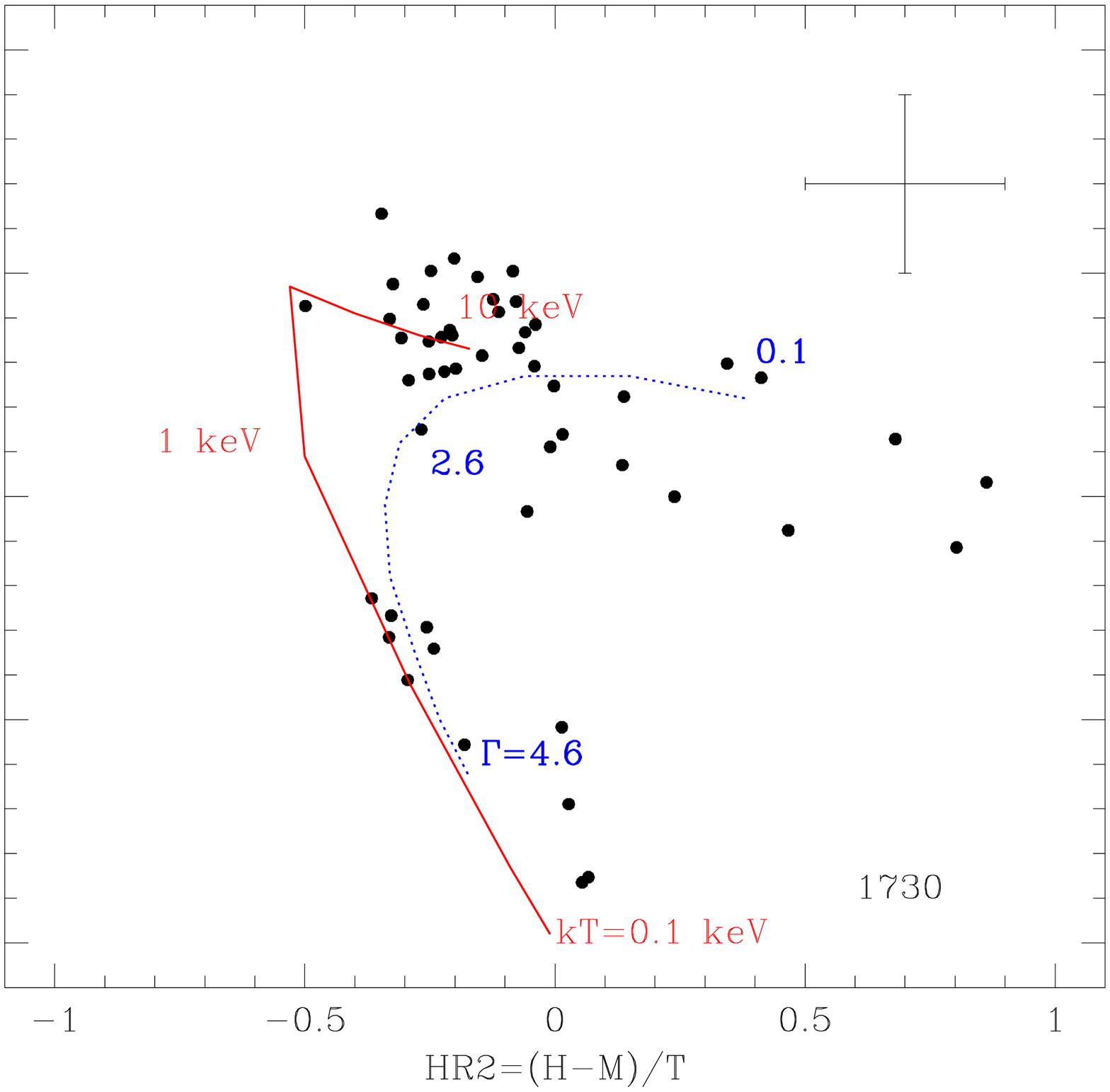}}
  \resizebox{0.3\hsize}{!}{\includegraphics{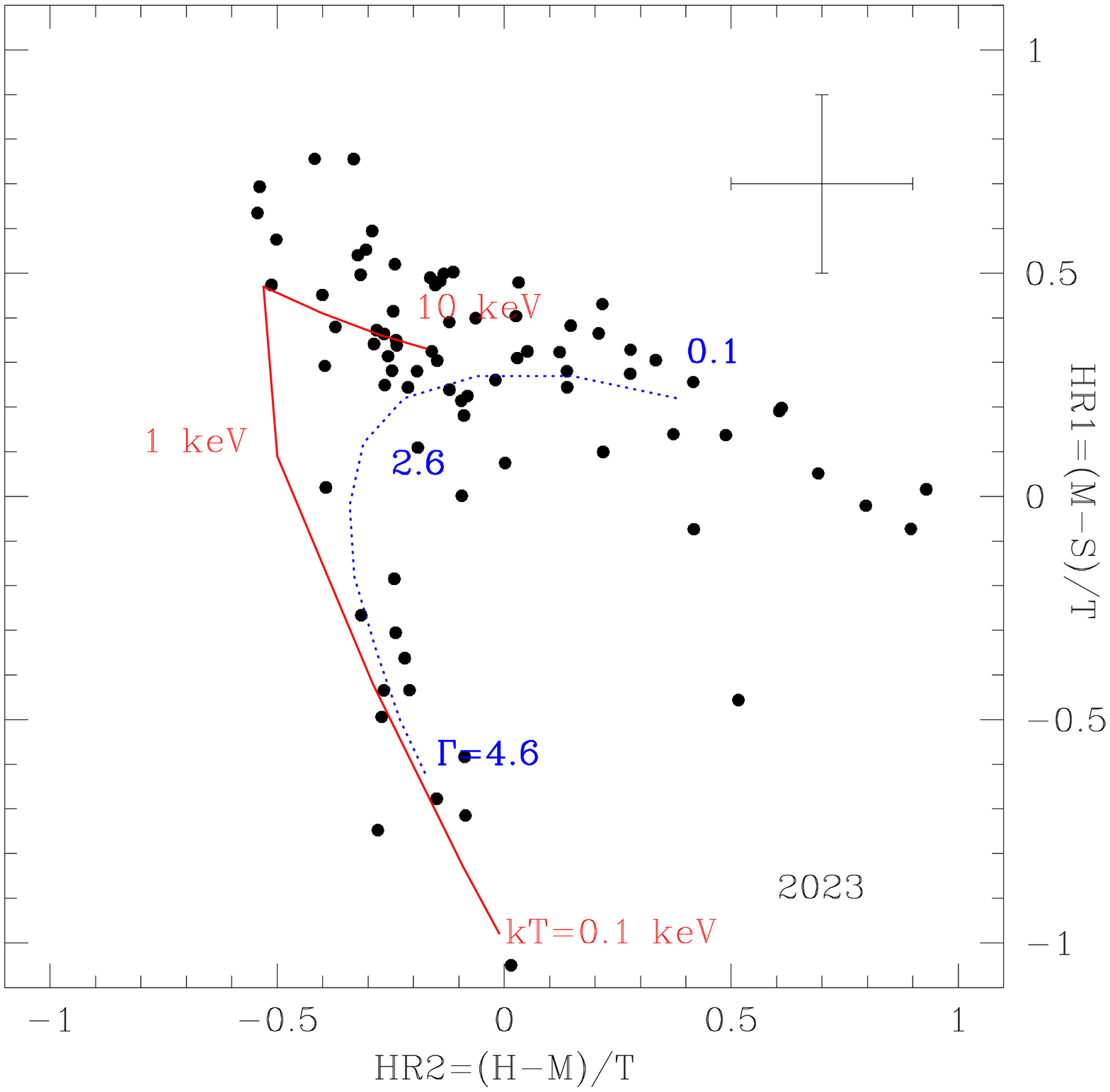}}
\end{center}
  \caption{Hardness ratio diagrams for the three
\chandra~observations. The solid line is a model path for a black body 
spectrum, the dotted line a model for a power law, both with Galactic
absorption and taking into account different responses of the main
CCDs in the observations. The CCD response is not a strong
effect. Increasing absorption moves the lines to the upper left
corner. The cross represents a typical error bar for a source with
35--40 counts.}
  \label{fig:color1}
\efige

In Sec. \ref{sec:disc} we will use optical identification together with
colors to investigate the color relations of different source types in
more detail.

\section{Luminosity function}
\label{sec:lf}

Since the \chandra~observations have different exposure times and cover 
somewhat different areas, we present here the luminosity functions for 
each observation separately. Fig.\ref{fig:lumi1} shows the observed
luminosity functions. The black lines show the sources
detected in the whole band (0.3--8.0 keV), the dotted line sources
detected in the soft band (0.3--2.0 keV), and the dashed line sources
detected in the hard band (2.0--8.0 keV).

The count to luminosity conversion was established based on the fit of 
source spectra with more than 100 counts with XSPEC 11.3.0. The
typical spectrum for the fainter sources in this range, $\sim$100--200
counts, is an absorbed power law with Galactic absorption, $6\times
10^{20}$ cm$^{-2}$, and a photon index of $\sim$2.0. This spectrum
roughly describes an X-ray binary spectrum, the sources of particular
interest. These values are assumed for fainter sources for the
conversion from counts to flux. The values quoted in the following for
sources with more than 100 counts are the fitted observed values. As a
rule of thumb 100 counts correspond to a luminosity of $1-2\times
10^{36}$ erg s$^{-1}$. The spectra of the brighter sources are roughly
consistent with such a spectrum. The details of spectral fitting will
be discussed in a separate paper.

\bfige
\begin{center}
  \resizebox{0.3\hsize}{!}{\includegraphics{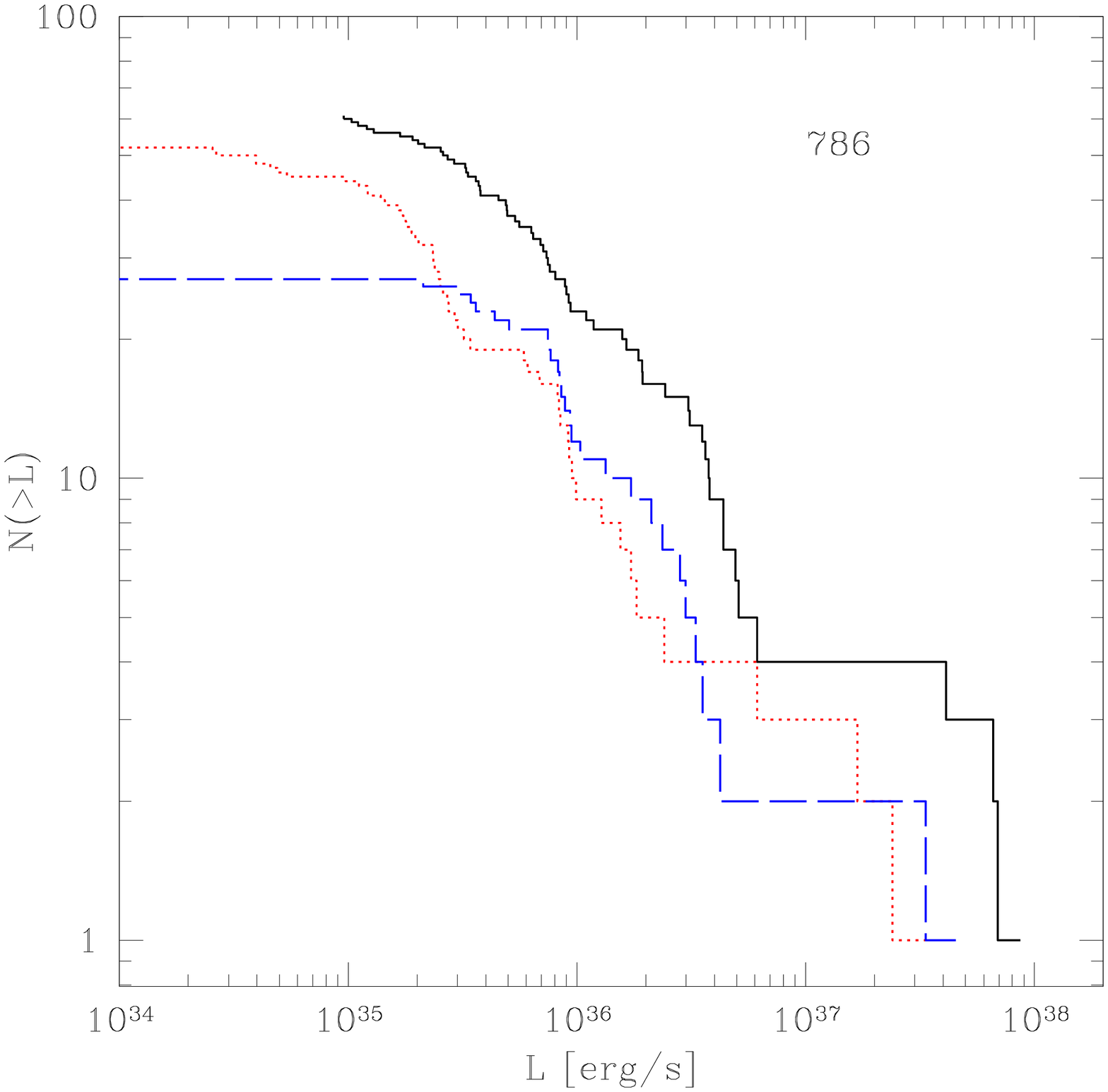}}
  \resizebox{0.3\hsize}{!}{\includegraphics{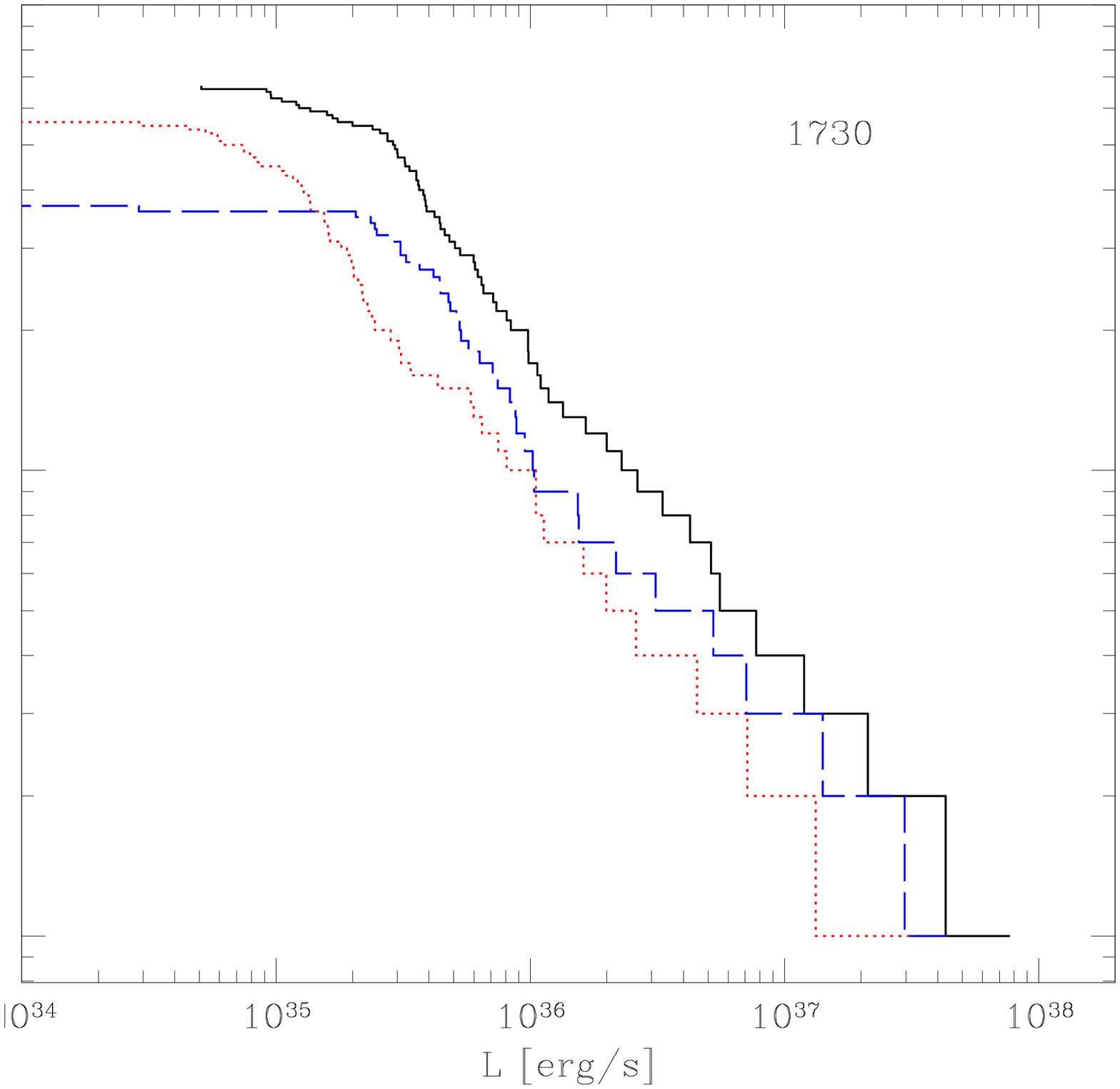}}
  \resizebox{0.3\hsize}{!}{\includegraphics{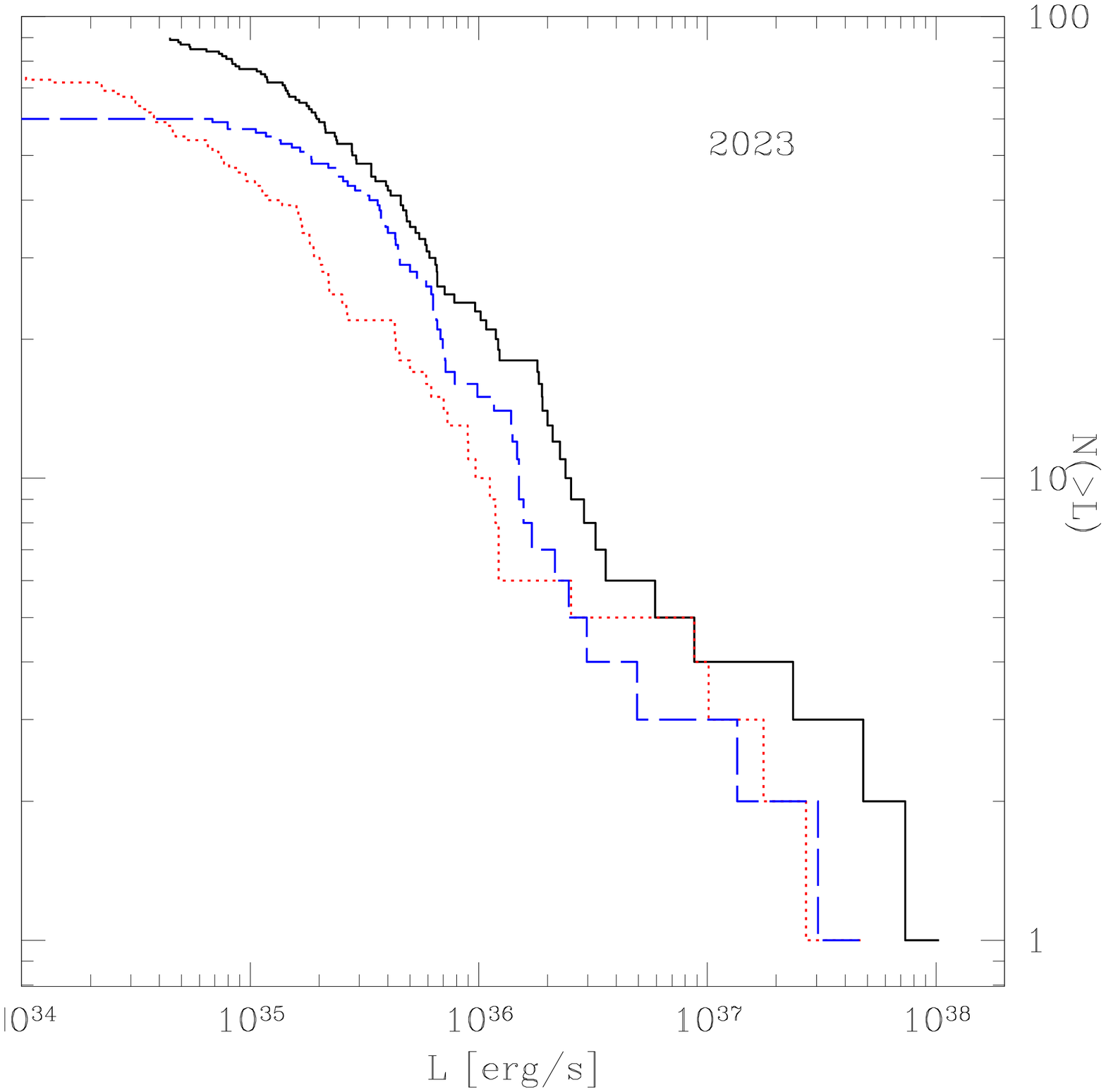}}
\end{center}
  \caption{Luminosity functions for the three \chandra~
observations. The black lines show the sources detected in the whole
band (0.3--8.0 keV), the dotted line sources detected in the soft band
(0.3--2.0 keV), and the dashed line sources detected in the hard band
(2.0--8.0 keV).}
  \label{fig:lumi1}
\efige

In Fig.\ref{fig:lumi2} we present the luminosity functions in the
full band after correcting for incompleteness by applying the
procedure outlined in \citet{kim:03}. A comparison with
Fig.\ref{fig:lumi1} shows the importance of this correction for
obtaining the actual shape of the luminosity function.
\bfige
\begin{center}
  \resizebox{0.3\hsize}{!}{\includegraphics{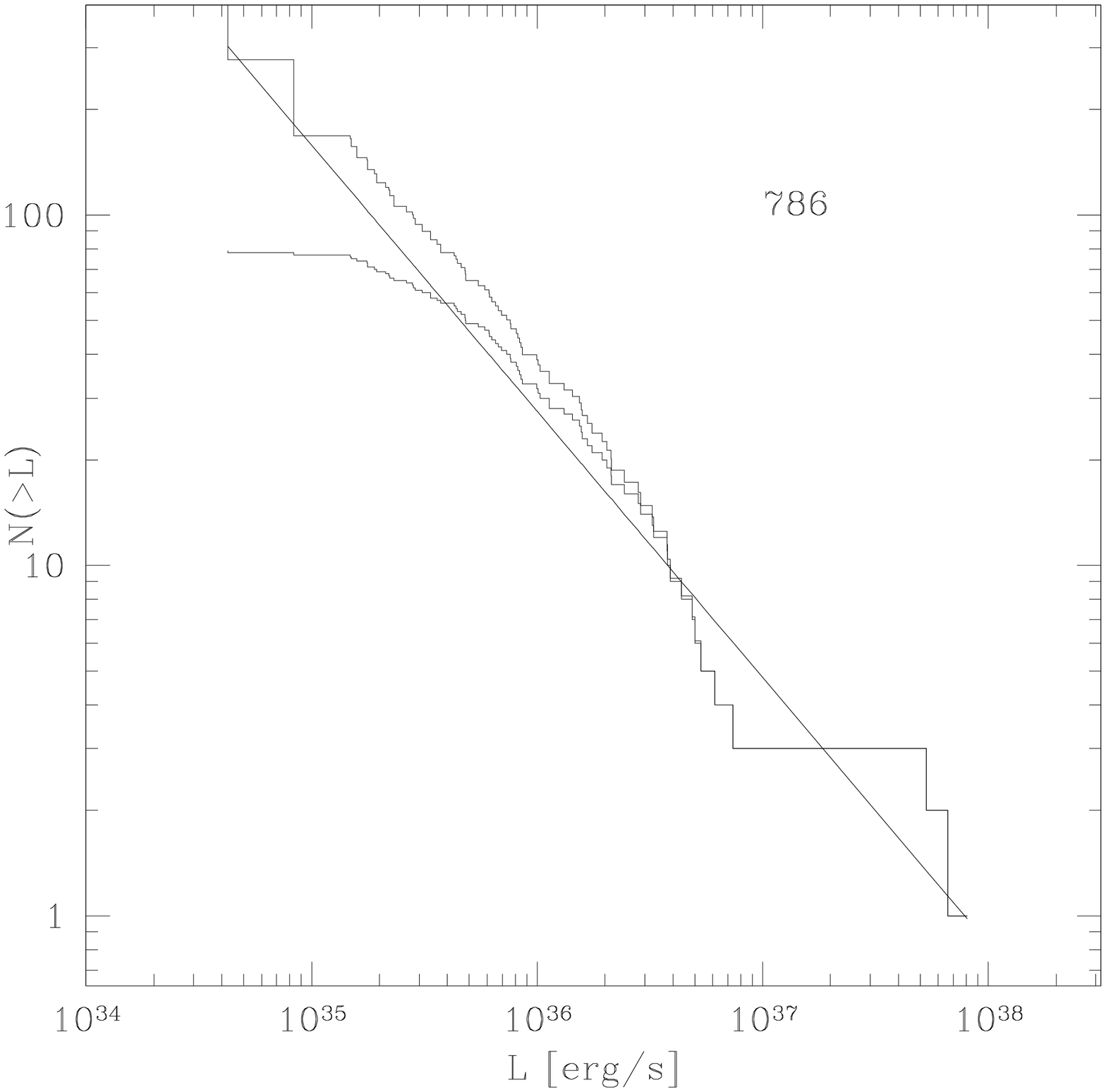}}
  \resizebox{0.3\hsize}{!}{\includegraphics{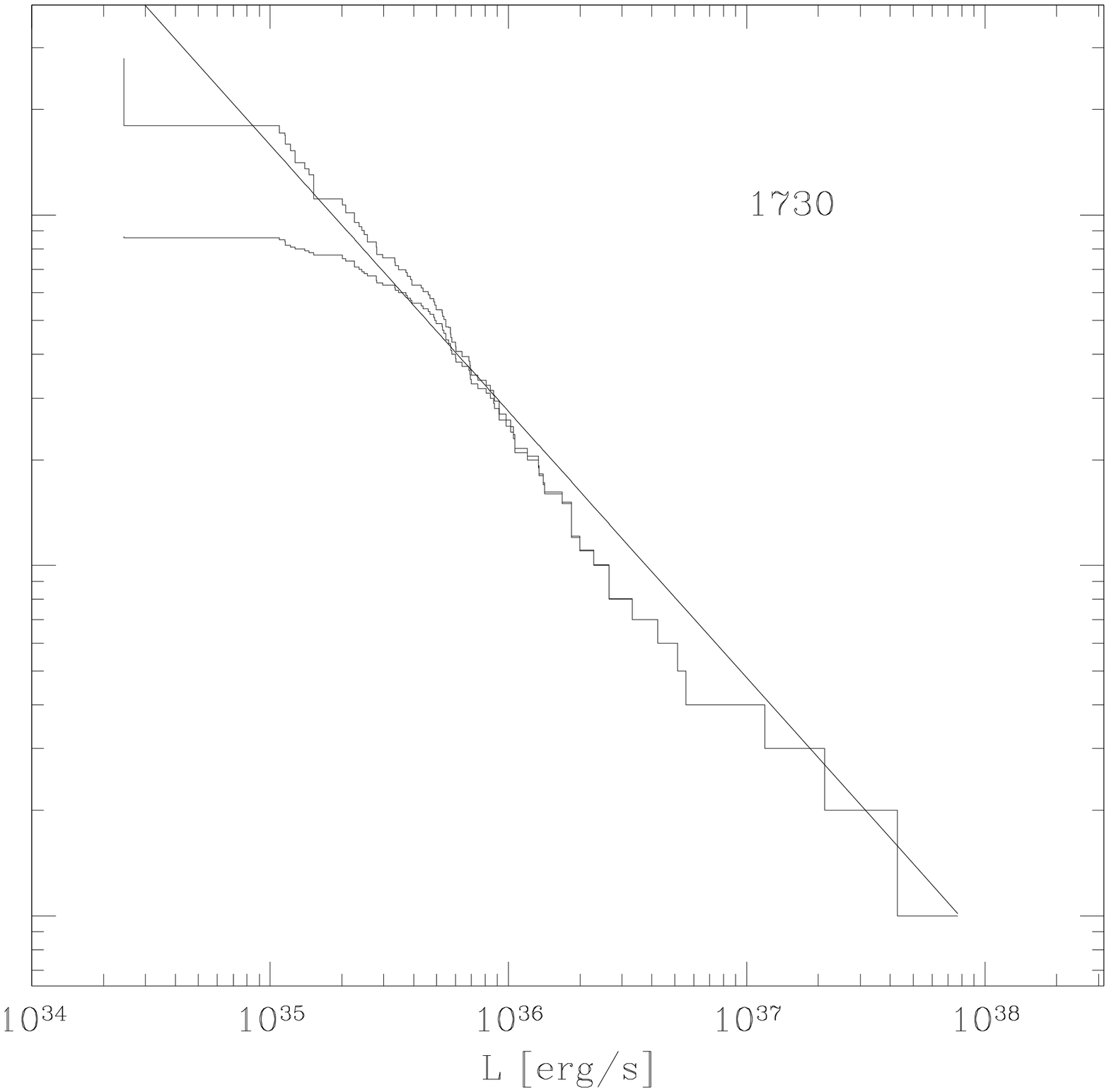}}
  \resizebox{0.3\hsize}{!}{\includegraphics{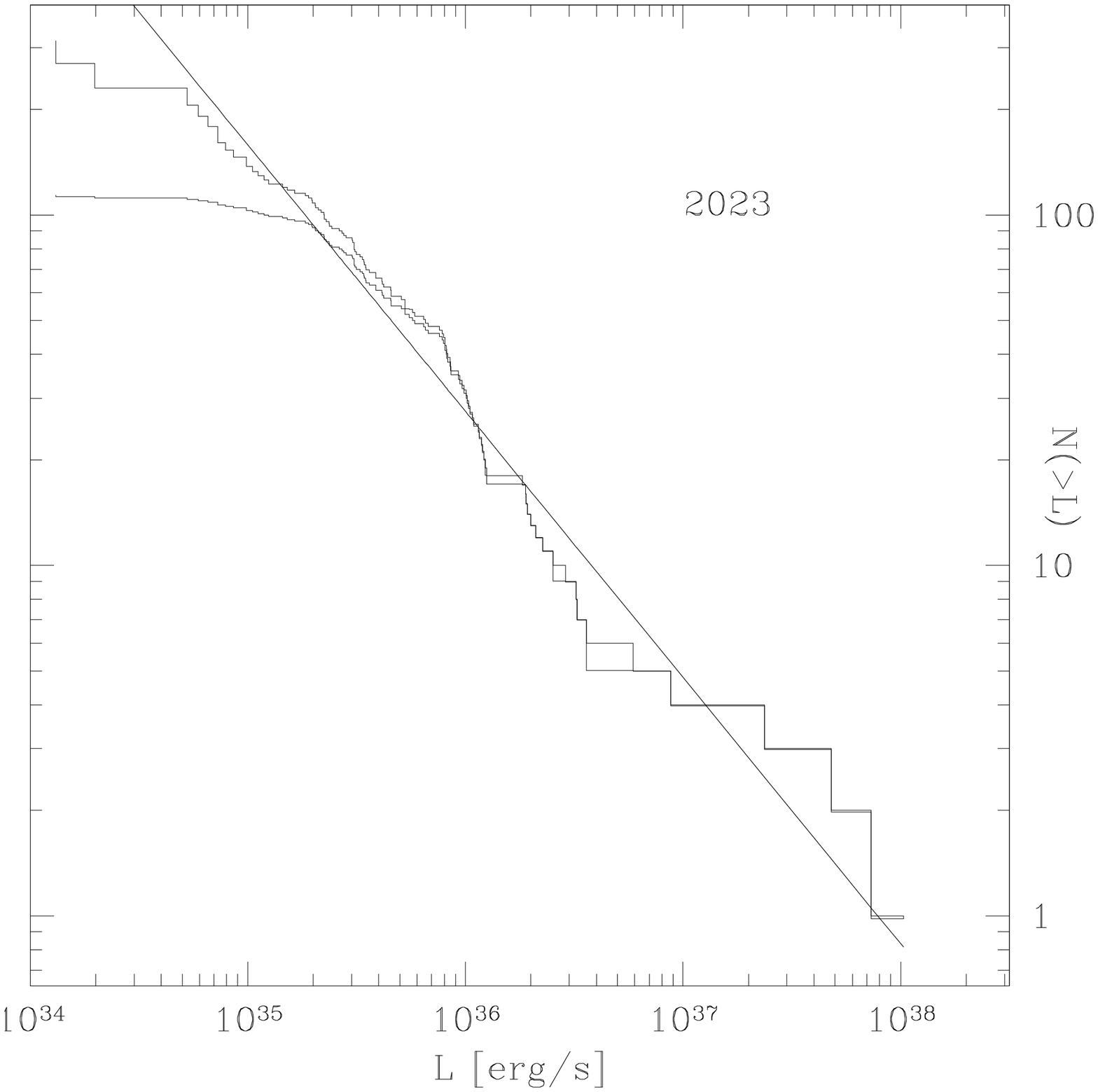}}
\end{center}
  \caption{Corrected and uncorrected luminosity functions for the
three \chandra~ observations. The histograms show the sources detected
in the whole band (0.3--8.0 keV). The solid line (the same in each
panel) is the best fit power law for the luminosity function of ObsId
1730. The scale is the same in all panels.}
  \label{fig:lumi2}
\efige

A Kolmogorov-Smirnov test shows that the luminosity functions have a
probability higher than 95\% to be drawn from the same distribution.
The slopes of the luminosity functions are relatively steep for an
actively star forming galaxy. A Maximum-Likelihood fit
gives slopes for cumulative luminosity function for
the observations 786, 1730, and 2023 of 0.78$\pm$0.17, 0.76$\pm$0.14,
and 0.74$\pm$0.13, respectively. The effect of interloper
contamination on these slopes is discussed in Sec. \ref{sec:disc_lf}.

\section{Counterparts}

In order to gain more information about the nature of the X-ray
sources we cross-correlate our sources with a number of different
catalogs available for M33. Given the proximity of M33 and the
resulting good resolution, there is a large amount of data in many
wavebands. We used the SIMBAD database to find catalogs
pertaining to M33, and we also cross-correlated the positions of
individual \chandra~sources against the whole SIMBAD database. We
discuss below the results of these comparisons with X-ray catalogs and 
multi-wavelength databases separately.

We cross-correlated our source list with the following
catalogs: The XMM-Newton survey of M33 \citep{pietsch:04}, the ROSAT
source catalog \citep{haberl:01}, the catalog of SNR by
\citet{gordon:99}, HII regions from \citet{hodge:99}, giant molecular
clouds from \citet{engargiola:03}, globular clusters from
\citet{mochejska:98}, star clusters from \citet{chandar:99} and
\citet{chandar:01}, blue and red supergiants from \citet{ivanov:93},
red supergiants from \citet{massey:98a}, Wolf-Rayet stars from
\citet{massey:98b}, UV bright stars from \citet{massey:96}, hot stars
from \citet{massey:95}, OB associations from \citet{regan:93},
planetary nebulae from \citet{magrini:01}, H$\alpha$ emitters from
\citet{calzetti:95}, and the DIRECT star catalog from
\citet{macri:01}.

The cross-correlation is done using the uncertainty of the \chandra~
source position, computed according to \citet{kim:04a}. In addition
the source uncertainty of catalog positions was quadratically added to
the \chandra~positional uncertainty. If no such number is available in the
catalogs, an uncertainty of 1 arcsec was added. If there are more than 
one source in the error circle we make no further assumption about the
likelihood of any of the sources {\it within} the error circle being a
real counterpart of the X-ray source.

\subsection{X-ray counterparts}

M33 has been observed with various X-ray missions, most notably ROSAT
and recently XMM-Newton. \citet{haberl:01} have analyzed all available
ROSAT observations and extracted a total of 184 sources. The
ROSAT observations cover a much larger area of M33 than our \chandra~
observations. Also note that the ROSAT (0.1--2.4 keV) and
\chandra~energy bands (0.3--8.0 keV) are different. Only 115 ROSAT
sources are in the field of view of the
\chandra~observations. Of these 115 sources, we detect 50 with
\chandra: Twelve of these are supernova remnants (SNR) as classified
by \citet{haberl:01}, three supersoft sources (SSS), two X-ray binaries
(XRB), including M33 X-7 \citep{peres:89} the first such
identification, and two foreground stars. Of the ROSAT detections 20
sources are far off-axis ($\ga$ 8 arcmin) in any \chandra~observation,
and 22 are observable in only one observation. Table \ref{tab:rosat}
in the Appendix contains the matches of \chandra~and ROSAT
sources. Column (1) is the \chandra~source name, column (2) the
\chandra~positional uncertainty, column (3) RA and Dec of the ROSAT
source, column(4) the ROSAT positional uncertainty, column (5) the
ROSAT name, column (6) the HRI flux and the corresponding 1$\sigma$
error. The typical astrometric error for the ROSAT sources is $\sim$5
arcsec with a long tail to larger errors; 32 sources have positional
uncertainties larger than or equal to 20 arcsec.

The XMM-Newton observations of M33 were of relatively shallow
exposure, lasting generally between $\sim$10--15 ks \citep{pietsch:04}.
Their detection limit of $\sim1.4 \times 10^{-15}$ erg s$^{-1}$ cm$^{-2}$
for a power law with $\Gamma=1.7$ and $N_H=6\times10^{20}$ cm$^{-2}$
compares to our $\sim 3\times 10^{-16}$ erg s$^{-1}$ cm$^{-2}$.
Like the ROSAT observations, this XMM survey also 
covers a larger area than \chandra. Of the 408 sources detected by
XMM, 184 are in the \chandra~fields of view. Of these 184 sources 102
are detected with \chandra, including 13 SNRs, two SSSs, one XRB,
three foreground stars, four background AGN, and 35 
hard sources as defined by \citet{pietsch:04}. Table \ref{tab:xmm} in
the Appendix contains the \chandra~and XMM matches. Column (1) is the
\chandra~source name, column (2) the \chandra~positional uncertainty,
column (3) RA and Dec of the XMM source position, column (4) XMM
energy flux and 1$\sigma$ error, column (5) the XMM X-ray
identification.

\subsection{Other wavebands}

The correlation with the radio data from \citet{gordon:99} yields 24
matches, 14 of which are SNRs or candidate SNRs based on their radio
spectral slope. One SNR matches two \chandra~sources. Five SNRs are
associated with HII regions. Nine matches are foreground/background
objects. The radius assigned for sources of the whole catalog is 3
arcsec. The optical radius of the SNRs is on average 4.5
arcsec. Taking the actual optical radius to match \chandra~sources
with selected SNRs in Table 4 of \citet{gordon:99} does not change the 
result for the SNR matches. We expect less than three chance coincidences
at 95\% for the whole catalog and less than two for the selected SNRs.

There is one association with a globular cluster in M33 from the sample of
\citet{mochejska:98}. The assumed radius of the clusters is 2
arcsec. The number of expected chance coincidences is less than
1. This source is also the only match with the samples of
\citet{chandar:99} and \citet{chandar:01} (also 2 arcsec radius). This
source does not match any other catalog, however its location in a
globular cluster suggests that it is either a bright CV or a low-mass
X-ray binary (LMXB). 

There are nine matches with blue stars in M33 from the sample of
\citet{ivanov:93} for seven \chandra~sources; two \chandra~sources
match two blue stars each. The accuracy for the blue star positions
is 1.5 arcsec \citep{ivanov:93}. None of the red stars in the
\citet{ivanov:93} sample matches any \chandra~source. Due to the
highly clustered nature of the blue supergiants it is not
straightforward to calculate the probability of chance
coincidences. However, it is possible that the majority of matches are
not real.

Due to the large number of objects, roughly 60000, a matching with
DIRECT sources yields ambiguous results. The limiting magnitude of the 
catalog is 23.6 in V. For off-axis angle sources ($\sim4$ arcmin)
where the \chandra~beam is large , there are numerous
associations. But for \chandra~sources closer to the aim point, and a
correspondingly smaller error radius ($\leq 1$arcsec), there are also
43 unique counterparts. However, because of the very high density of
DIRECT objects, most of them are likely chance coincidences.

Similarly to the DIRECT sources, OB stars from \citet{regan:93} have a
high density and give multiple matches. In total there 14 matches with
nine \chandra~sources, six of which are unique matches. The positional
accuracy of the OB stars is within 2 arcsec \citep{regan:93}. The high 
source density and overlapping source regions indicate that the
majority of matches are chance coincidences.

The HII region sample of \citet{hodge:99} gives three matches. As
radius for the HII regions we use the optical radius if given,
otherwise we use 1 arcsec. Despite the small radius, the number of
expected chance coincidences is still three because of the large
number of HII regions ($\sim 1200$).

The UV stars in M33 in the sample of \citet{massey:96} yield six
matches (uncertainty 1.4 arcsec, 2 chance coincidences expected). One
of the sources is identified with a B1 supergiant, another with a LBV
candidate, and another one with an HII region. One of the
\chandra~sources is also the only match with the Wolf-Rayet star
sample of \citet{massey:98b}. There is also only one match with the
catalog of hot stars of \citet{massey:95}. For the two later catalogs
an uncertainty of 1 arcsec was assumed and 1 chance coincidence is
expected in each.

\section{Discussion}
\label{sec:disc}

\subsection{Source types}
\label{sec:disc1}

\bfig
  \resizebox{0.48\hsize}{!}{\includegraphics{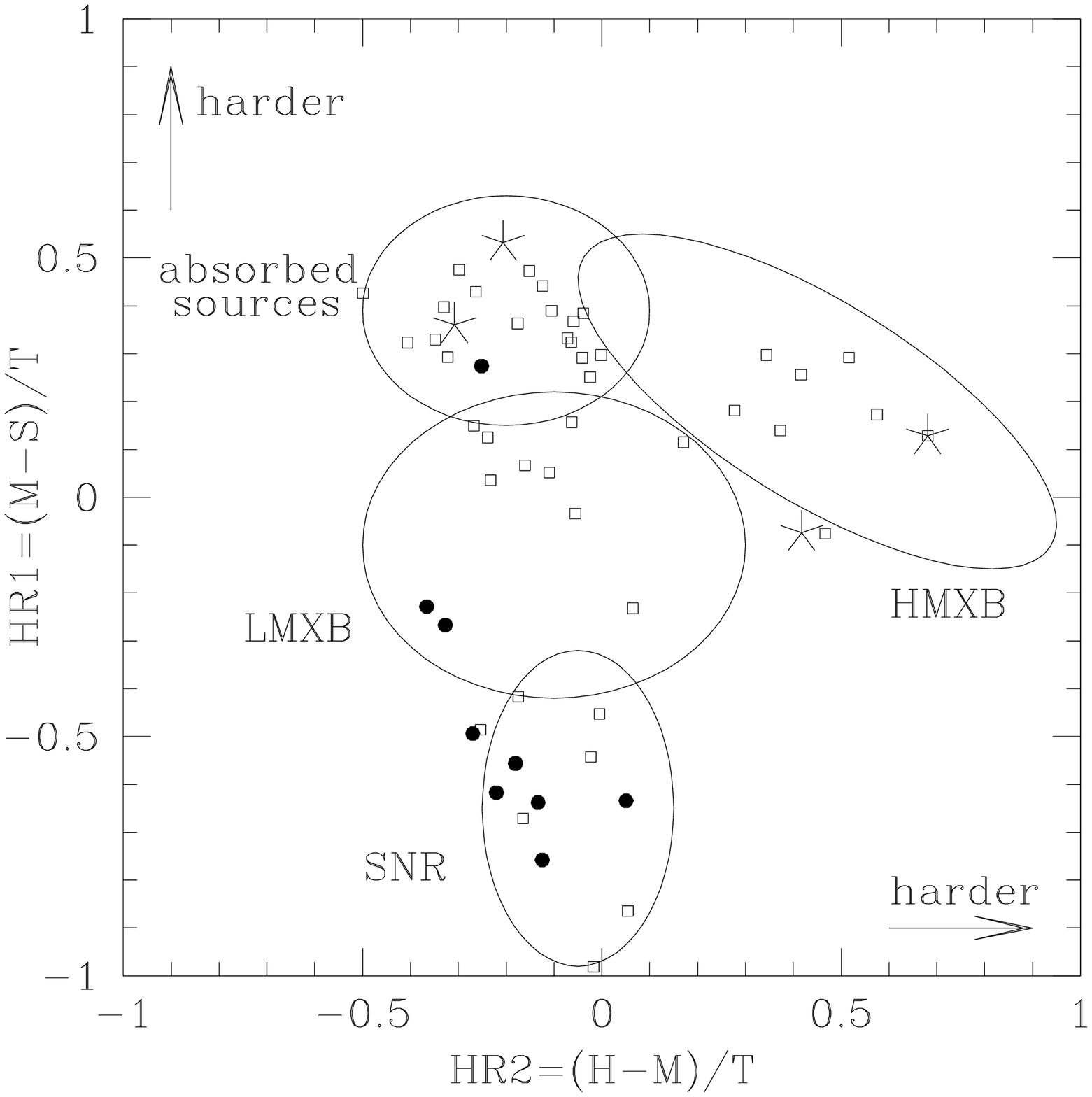}}
  \caption{Hardness--ratio diagram of sources with
  identifications. Filled circles--SNR, stars--background/foreground
  objects, open squares--''blue'' counterparts. The elliptical areas
  indicate the expected location of various source types. The areas
  are taken from \citet{prestwich:03}. Only sources with no upper
  limits in any flux band are shown. For more details see
  Sec. \ref{sec:disc1}.} 
  \label{fig:col_comp}
\efig

Due to the comparable number of expected background sources and sources
associated with M33 in our dataset it is important to distinguish the
different source types. Only by identifying the X-ray sources is it
possible to learn more about the true source population of M33. As
a first step we compare X-ray colors with the expected location of
different source types in the hardness--ratio diagram. Again sources
with upper limits in any band are disregarded. In
Fig.\ref{fig:col_comp} we show different source types, derived from
identifications in other wavelengths: filled circles show
supernova remnants; stars are background and foreground objects; open
squares are objects with a ``blue'' counterpart, blue meaning either
an H$_{\alpha}$, UV, U or B band counterpart detected in M33. As can
be easily seen from the picture, SNRs do populate the expected locus
in the hardness--ratio diagram based on the prescription of
\citet{prestwich:03}. There is only one exception of an SNR from the
Gordon catalog which has a harder X-ray spectrum than expected of a
thermal SNR, and might either be a plerionic SNR or an SNR harboring
an X-ray binary. 

Comparing the X-ray hardness ratio diagram with the regions
\citet{prestwich:03} identified with certain source types in
Fig.\ref{fig:col_comp} shows that the region that covers LMXBs,
around (-0.3,0) in HR2--HR1, is only sparsely populated whereas all
other regions are well populated with sources. The distribution of
points in the color diagram, and in particular the avoidance of the
LMXB locus is reinforced if we consider Fig. \ref{fig:color1} which
contains both identified and unidentified sources. AGN and pulsar-like
X-ray binaries are not distinct in the hardness ratio diagram, both
being consistent with hard and/or possibly absorbed sources. Note that
very strongly absorbed sources ($N_H\ge 10^{22}$ cm$^{-2}$) would not
be plotted because they would not be detected in the soft band if they 
were not very bright. This implies that LMXBs do not contribute
significantly to the X-ray source population in M33. This is
consistent with the estimate of its stellar mass and star formation
rate which suggests a prevalent young stellar population.

Although the regions shown in Fig.\ref{fig:col_comp} do not represent
a quantitative expectation value 
for different source locations, we can nevertheless quantify our
expectations for the number of HMXBs and LMXBs from the universal
luminosity functions for HMXBs \citep{grimm:03} and LMXBs
\citep{gilfanov:04}. We take the star formation rate of M33 to be
$\sim$0.3 \md yr$^{-1}$ \citep{hippelein:03}, and the K band
magnitude to be $m_K = 4.1$ \citep{jarrett:03}. Using eq. (22) of
\citet{grimm:03} and eq. (21) with the normalization for late-type
galaxies of \citet{gilfanov:04}, we expect to observe $\sim6-7$ HMXBs
and $\sim3$ LMXBs above $10^{37}$ erg s$^{-1}$. At $10^{37}$ erg
s$^{-1}$ we expect no strong contamination by background
sources. These numbers are in good agreement with the observed number
of 6 sources above $10^{37}$ erg s$^{-1}$. Although the
\chandra~observations do not cover M33 in its entirety they cover more
than 80\% of M33. The expected total luminosity is
$\sim3.5\times10^{38}$ erg s$^{-1}$ for HMXBs and $\sim10^{38}$ erg
s$^{-1}$ for LMXBs. These values agree very well with the observed
total X-ray luminosity of point sources (excluding the nucleus) of
$\sim3-4\times10^{38}$ erg s$^{-1}$. Using the formulae of
\citet{colbert:04} for the relation of total X-ray luminosities with
K-band luminosity, FIR+UV luminosities, as well as with stellar mass
and SFR give basically identical results. An exception is the
X-ray--stellar mass relation, which predicts an LMXB X-ray luminosity
of $\sim1.2\times10^{39}$ erg s$^{-1}$, an order of magnitude larger
than predicted by other relations, and 3--4 times larger than the
oberved luminosity.

\subsection{Luminosity function}
\label{sec:disc_lf}
Based on spectrophotometry and comparison with theoretical SEDs M33 is
expected to be dominated by a young stellar population
\citep{li:04}, in agreement with the source colors as discussed in
Sec. \ref{sec:disc1}. Therefore most X-ray sources in M33 should be
high-mass X-ray binaries (HMXB). We would then expect the luminosity
function of the M33 sources to be similar to the luminosity function
of HMXBs in the Milky Way. Fig.\ref{fig:hmxb_lf} shows the corrected
luminosity functions of M33 and the luminosity functions of HMXBs in
the Milky Way (thick solid histogram) and in the SMC (dotted
histogram). The luminosity functions differ slightly in shape but have
a similar slope within the errors. The SMC luminosity function seems
to be somewhat flatter but there is a large contribution from the four
brightest sources in the SMC. On the other hand the M33 luminosity
functions obtained from Maximum-Likelihood fits are steeper
($\sim$0.76$\pm0.14$) than the Milky Way HMXB luminosity function
(0.64$\pm$0.15) though still consistent within the errors.

These values for the slope of the M33 luminosity functions lie in
between what is found for star forming galaxies, $\alpha \sim
0.5-0.6$ \citep{kilgard:04}, and elliptical galaxies, $\alpha \sim 1$
\citep{kim:04c}. The reason for the steep luminosity function is most
likely the expected large number of background objects ($\sim$50\%)
which will steepen the luminosity function due to their steep
number--flux relation \citep{alexander:03}. A significant contribution
of LMXBs is unlikely since the stellar mass of M33 is roughly one
tenth of the Milky Way and therefore the number of LMXBs are expected
to be similarly small ($\sim10$ above $10^{36}$ erg s$^{-1}$ in all of
M33). Moreover, at low luminosities ($L_X < 10^{37}$ erg s$^{-1}$) the
LMXB luminosity function flattens as well \citep{grimm:02,kong:02}.

We compared our number--flux relation with the 2Ms CDF-N source
catalog of \citet{alexander:03} in the full band\footnote{The data
were obtained from Neil Brandt's web page
http://www.astro.psu.edu/users/niel/hdf/hdf-chandra.html}.
In order to compare the CDF-N data with our analysis of M33 we take
into account different assumptions about spectral shape, different fields
of view and different column densities. \citet{alexander:03} assume a
power law spectrum with a photon index of $\Gamma = 1.4$, a Galactic
absorption column density of $N_H = 1.6\times10^{20}$ cm$^{-2}$, and
compute the flux in the energy range from 0.5--8.0 keV. Converting the 
fluxes to our values of $\Gamma = 2$ and $N_H = 6\times10^{20}$
cm$^{-2}$ in the energy range from 0.3--8.0 keV results in a
correction factor of $\sim$0.5 for the CDF-N data. Moreover, the
different fields of view used in the CFD-N analysis and our analysis
(note we correct our observations individually) yields another factor
of $\sim$0.6 with respect to the CDF-N data for the number of
sources. The first factor obviously depends only on assumptions about
spectral shape and energy band used. The second factor, however, is
subject to uncertainties arising from the (non-)existence of cosmic
variance. This background can significantly affect the normalization of 
the background Log(N)--Log(S).

We find that after subtracting the corrected CDF-N number--flux
relation from the M33 ones the slope flattens to $\sim 0.5-0.6$. Note
that this is an estimate of slope, not a fit. The slopes are flatter
for higher background normalizations. However the range is relatively
robust for reasonable assumptions about the normalization of the
background Log(N)--Log(S) arising from cosmic variance, i.e. changes
in normalization of a factor of 2--3. We also compare the soft band
(0.3--2.0 keV) luminosity function with ROSAT All-Sky Survey data from
fields adjacent to M33. Slope and normalization of the RASS
Log(N)--Log(S) are consistent with the values from the CDF-N in the
overlapping flux range. The solid line in Fig.\ref{fig:hmxb_lf} is an
approximation to the {\em intrinsic} M33 luminosity function with a
slope of 0.55, and a normalization fixed by the brightest sources. A
more detailed study of luminosity functions and background
Log(N)--Log(S) is necessary to quantify these results further.

Fig.\ref{fig:m33_m31} shows the luminosity function of M33
overlayed with the fitted luminosity functions for different regions
of M31 \citep{kong:03}. The dashed line is the luminosity function for
a region in the outer parts of M31 and has a slope of 1.1, the
dash-dotted line is for a region in the disk of M31, and the dotted
line is for a region including parts of the star forming ring around
the bulge of M31. The solid line is again the approximation to the
{\em intrinsic} M33 luminosity function. The dash-dotted and dotted
lines, dominated by young stellar populations, resemble the luminosity
functions of M33 most closely.
\bfig
  \resizebox{0.48\hsize}{!}{\includegraphics{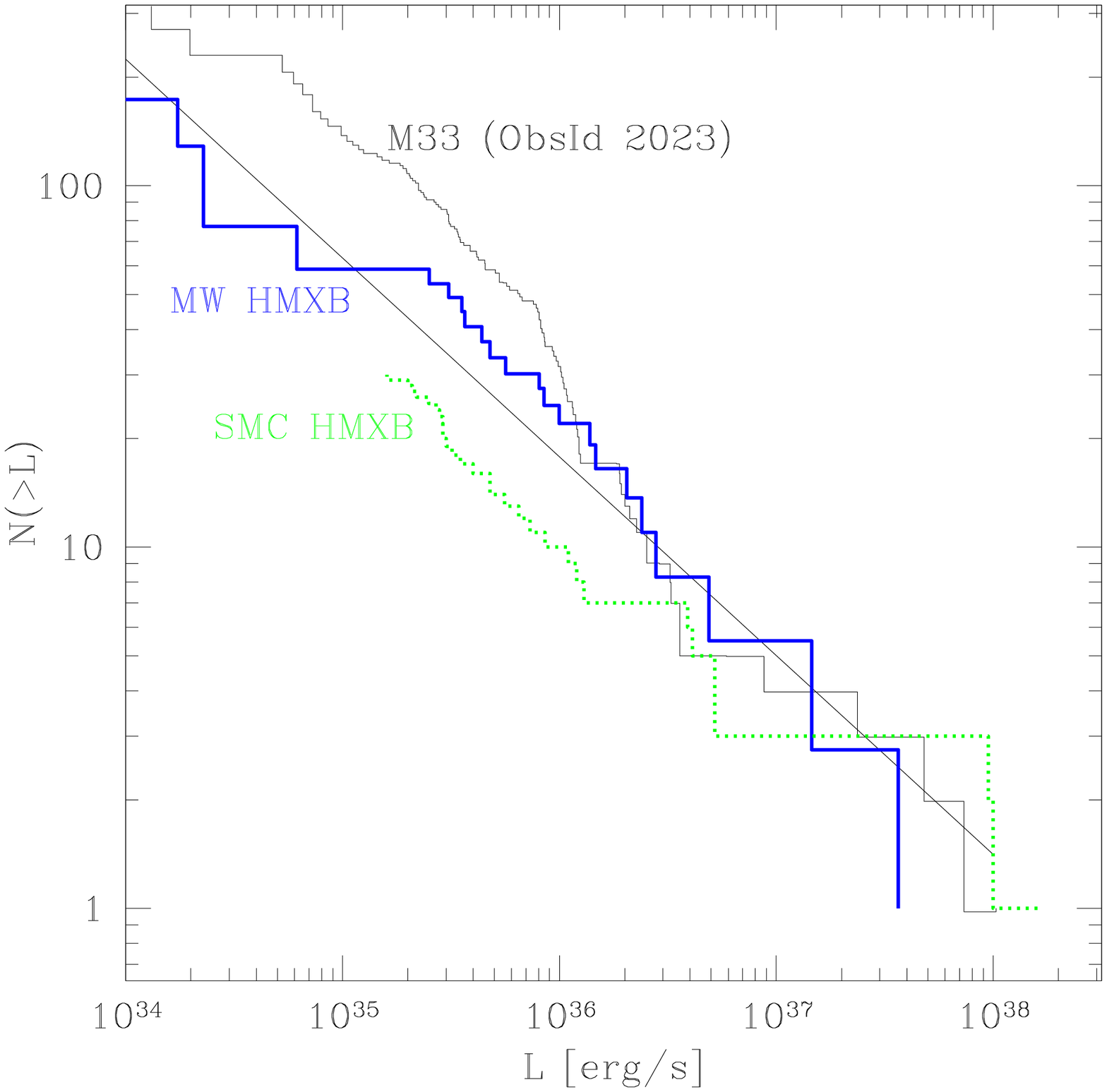}}
  \caption{Luminosity functions of X-ray sources in M33 from ObsId
   2023 (solid histograms) compared with the HMXB luminosity function
   of the Milky Way (thick solid histogram) \citep{grimm:02} and X-ray 
   binary candidates in the SMC (thick dotted histogram)
   \citep{yokogawa:00}. The solid line is a representation of the
   expected intrinsic luminosity function of M33 after subtraction of
   background AGN with a cumulative slope of 0.55. The normalization
   of the expected intrinsic luminosity function of M33 is defined by
   the high luminosity end of the luminosity function of ObsId 2023.}
  \label{fig:hmxb_lf}
\efig
\bfig
  \resizebox{0.48\hsize}{!}{\includegraphics{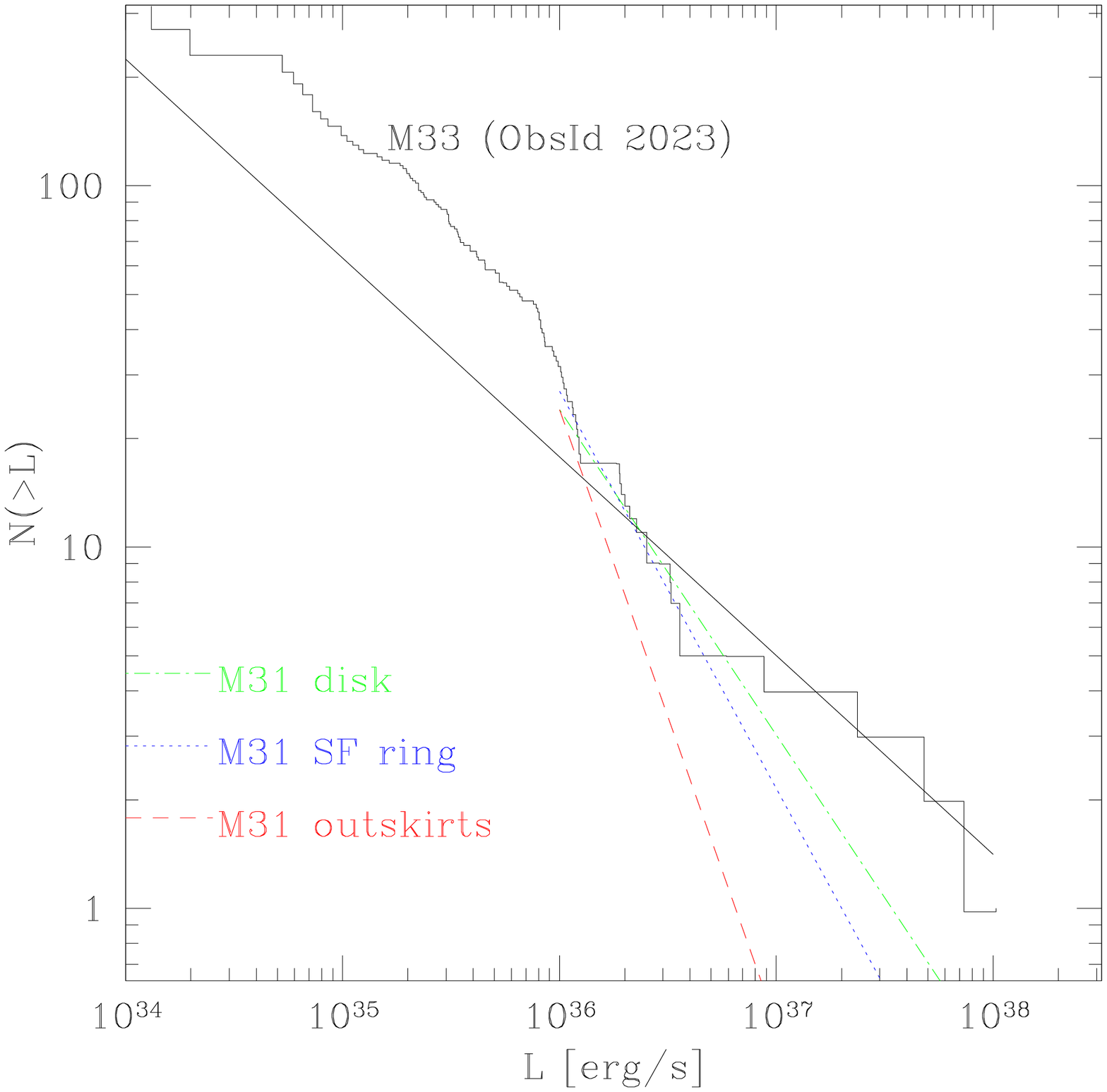}}
  \caption{Luminosity functions of X-ray sources in M33 from ObsId
   2023 compared with luminosity function fits from regions in M31 by
   \citet{kong:03}. The dashed line is the luminosity function (slope
   of 1.1) for a region in the outer parts of M31, the dash-dotted
   line is for a region in the disk of M31, and the dotted line is for
   a region including parts of the star forming ring around the bulge
   of M31. The solid line is a representation of the expected 
   intrinsic luminosity function of M33 after subtraction of
   background AGN with a cumulative slope of 0.55 and a normalization
   defined by the high luminosity end of the observed luminosity
   function.}
  \label{fig:m33_m31}
\efig

The similarity of the {\em intrinsic} slopes in M33 and the shapes of
the luminosity function with HMXB luminosity functions in the Milky
Way and starburst galaxies shows, that the X-ray source population in
M33 is dominated by HMXBs and other young X-ray sources. It also
agrees with the assumption that LMXBs do not contribute significantly
to the X-ray binary population in M33.

\subsection{Supernova remnants}
\bfig
  \resizebox{0.48\hsize}{!}{\includegraphics{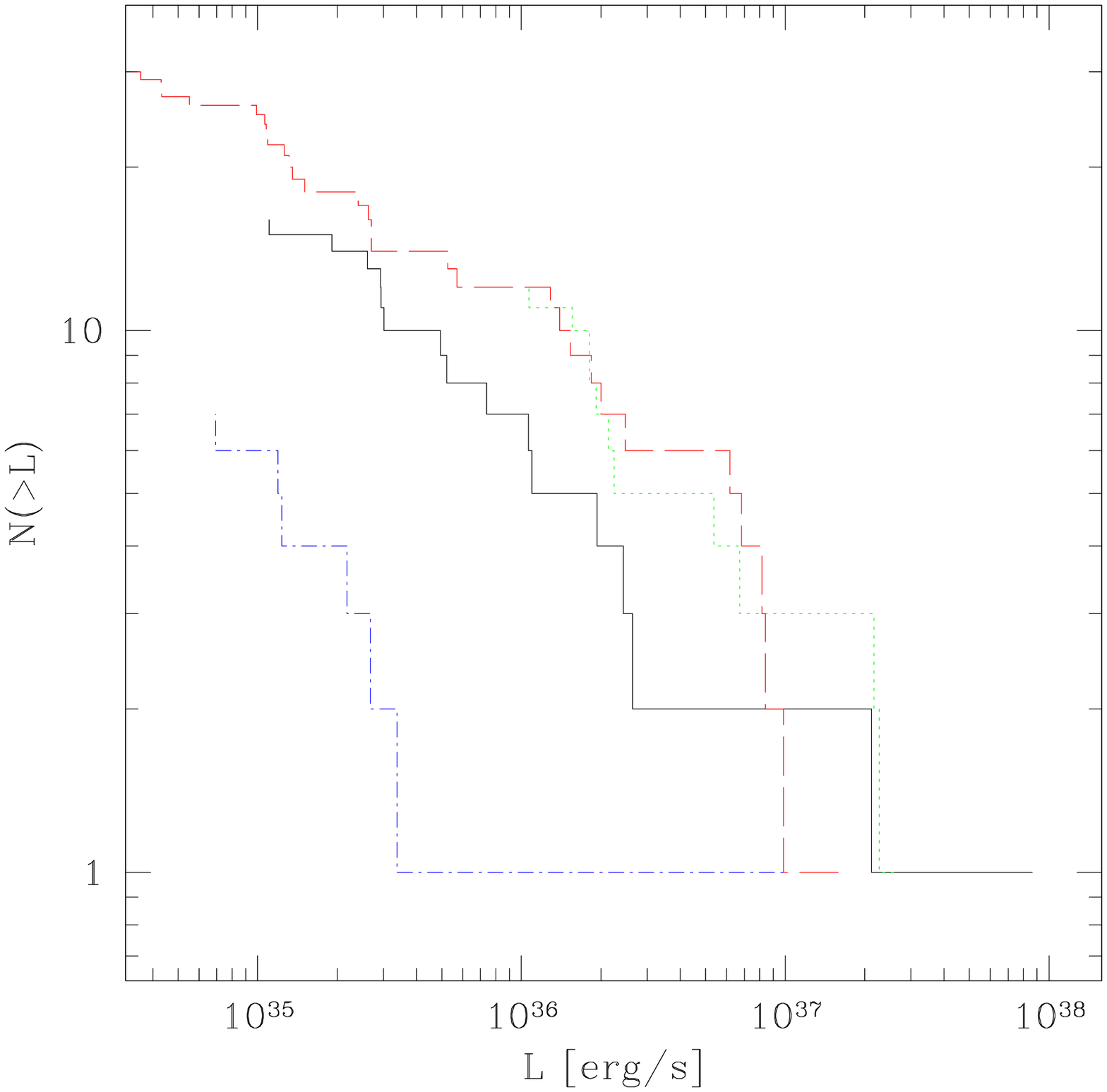}}
  \caption{Observed luminosity function of radio SNR in M33 combined
from the three \chandra~observations. For comparison we show the
observed luminosity of the Magellanic Clouds (dot-dashed
histogram--SMC, dashed histogram--LMC) from \citet{sasaki:00a} and
\citet{sasaki:00b}, and of M31 (dotted histogram) from \citet{supper:01}.}
  \label{fig:snr_lf}
\efig

Supernova remnants are another source type associated with young
stellar populations. The X-ray luminosity distribution of 16 SNRs
detected both in radio \citep{gordon:99} and with \chandra~is
shown in Fig.\ref{fig:snr_lf}. Due to the construction of the
correction of the intensity distribution by \citet{kim:04a} it is not
possible to correct the luminosity function of a subset of
sources. Therefore the SNR luminosity function is not corrected for
incompleteness.

For comparison we show the luminosity function of SNRs in M33, and in
the Magellanic Clouds and M31 obtained from ROSAT data in
Fig.\ref{fig:snr_lf}. The data for M31 (dotted histogram) are taken
from observations by \citet{supper:01}, the data on the
Magellanic Clouds (dashed histogram for LMC, dot-dashed histogram for
SMC) are from \citet{sasaki:00a} and \citet{sasaki:00b}. The data are
not corrected for the different energy bands of \chandra~and
ROSAT. For thermal supernova remnants most of the flux is in the soft
band, so the extended \chandra~response towards hard X-rays will not
significantly change the luminosity for these sources. Also different
assumptions about spectral shape do not strongly affect the
luminosities in this energy range. Although numerous SNRs are known
in the Milky Way, their poorly determined distances
and non-uniform sampling do now allow a comprehensive comparison with
the population with extragalactic SNRs.

Although there are relatively few sources, SNRs in the SMC are
underluminous compared with SNRs in the other galaxies. This has been
noted already by \citet{haberl:01} and the authors attribute this to
to higher metal abundances in the LMC. Although the luminosities in
M33, M31, and the LMC are comparable, a two sample Kolmogorov-Smirnov
test gives a probability of only $\sim$1\% for the luminosity
functions of M31 and M33, and the Magellanic Clouds, to have the same
underlying distribution. However, for M33 and the Magellanic Clouds
the probability increases to$\sim$10\%. And for LMC and SMC the
probability increases to over 60\%. Even considering the different
incompleteness limits and instrumental effects (M31 -- PSPC, the
Magellanic Clouds -- HRI) there nevertheless might be a difference in
the SNR population in star forming and older galaxies. For a complete
and more concise discussion of the SNRs in M33, refer to
\citet{ghavamian:05} who investigate the properties of the SNRs
individually.

\section{Conclusion}
The sensitivity of the \chandra~observations of M33, and the wealth of
multi-wavelength data for this galaxy does allow a uniquely detailed
study of the X-ray point source population.

Here we present the source list, counterparts, X-ray colors, and X-ray
luminosity functions of sources in M33 observed with \chandra. In
total 261 individual sources are detected in 3 partly overlapping
observations down to a flux of $3\times 10^{-16}$ erg s$^{-1}$
cm$^{-2}$, corresponding to a luminosity of $\sim 2 \times 10^{34}$
erg s$^{-1}$ at a distance of 840 kpc. The observations are centered
on the nucleus and the star forming region NGC 604. The luminosity
functions of the observations are statistically consistent with each
other. Because of the large area covered by the observation, $\sim$0.2 
square degrees, the number of background objects is large. Based on
CDF-N data we conclude that around 50\% of the detected sources are
background objects at low luminosities ($10^{34} \le L_X \le 5 \times
10^{35}$ erg s$^{-1}$).

Taking into account the contribution of background AGN to the
luminosity function of M33, the slope of the luminosity function,
cumulative $\sim 0.5-0.6$, is consistent with HMXB luminosity
functions in the Milky Way and the Magellanic Clouds. Color analysis
of the sources shows a lack of sources at the location populated by
low-mass X-ray binaries in the X-ray color-color diagram. Also
cross-correlations with catalogs at other wavelengths result
preferably in matches with object associated with young stellar
populations: e.g. SNRs, blue stars, HII regions. We therefore conclude
that the X-ray source population in M33 is dominated by a young
population, similar to other star forming galaxies. This is also in
agreement with expectations based on the stellar mass and star
formation rate of M33.

We also find that the shape of the luminosity function of SNRs in M33 is
more similar to the SNR luminosity functions in the Magellanic Clouds
\citep{sasaki:00a,sasaki:00b} than to the SNRs in M31
\citep{supper:01}. This might be related to abundance differences in
these galaxies \citep{haberl:01}.

The spectral properties, along with temporal properties, of sources
with a sufficient number of counts will be discussed in a following
paper.

\section{Acknowledgments}
This work has been supported by NASA grant GO2-3135X. The authors want
to thank Ralph Kraft for his program to compute upper limits, and
Wolfgang Pietsch for providing the XMM source list before
publication. We also thank the referee for constructive comments on
the paper.

\bibliographystyle{apj}
\bibliography{ms}

\appendix

\begin{deluxetable}{crrc@{\,}c@{\,}cc@{\,}c@{\,}cc@{\,}c@{\,}cr}
\tabletypesize{\tiny}
\rotate
\tablecaption{Source list ordered with increasing right ascension.\label{tab:sources}}
\tablewidth{0pt}
\tablehead{
\colhead{Source} & \colhead{Coord./Unc.} & \colhead{ObsId} & \multicolumn{6}{c}{Counts} &
\multicolumn{3}{c}{Flux[$10^{-7}$cts/s/cm$^2$]\tablenotemark{a}} & \colhead{L [erg/s]\tablenotemark{a}}\\
\colhead{CXO J}& \colhead{RA/Dec/["]} & \colhead{} & \colhead{0.3--8.0} & \colhead{0.3--2.0} & \colhead{2.0--8.0} & \colhead{0.3--1.0} & \colhead{1.0--2.1} & \colhead{2.1--8.0} & \colhead{0.3--8.0} & \colhead{0.3--2.0} & \colhead{2.0--8.0} & \colhead{0.3--8.0}
}
\startdata
013219.6+304153 & 01:32:19.668  & 1730 &    8.2(4.3)     &     $<$7.5      &    6.9(4.0)     &      ---     &      ---     &      ---     &   8.6(4.4)  &    $<$5.6   &   8.4(3.4)  &   1.54e+35\\
                & 30:41:53.57   &  786 &       ---       &       ---       &       ---       &      ---     &      ---     &      ---     &      ---    &      ---    &      ---    &     ---   \\
                & 27.0          & 2023 &       ---       &       ---       &       ---       &      ---     &      ---     &      ---     &      ---    &      ---    &      ---    &     ---   \\[0.1cm]
013239.7+303245 & 01:32:39.787  & 1730 & 11.0\tablenotemark{b}(3.3) &     $<$9.3      &    9.6(4.7)     &      ---     &      ---     &      ---     &   8.2\tablenotemark{b} &    $<$7.0   &   9.6(3.6)  &   1.47e+35\tablenotemark{b}\\
                & 30:32:45.78   &  786 &       ---       &       ---       &       ---       &      ---     &      ---     &      ---     &      ---    &      ---    &      ---    &     ---   \\
                & 12.7          & 2023 &       ---       &       ---       &       ---       &      ---     &      ---     &      ---     &      ---    &      ---    &      ---    &     ---   \\[0.1cm]
013253.4+303817 & 01:32:53.481  & 1730 &  969.6(32.8)    &  703.5(27.8)    &  237.5(17.1)    &  226.9(16.3) &  492.8(23.4) &  246.9(17.7) &   800(27)   &   560(21)   &   220(15)   &   1.43e+37\\
                & 30:38:17.07   &  786 &       ---       &       ---       &       ---       &      ---     &      ---     &      ---     &      ---    &      ---    &      ---    &     ---   \\
\enddata

\tablenotetext{a}{For sources with more than 100 counts, luminosity was
computed from spectral fit, otherwise $\Gamma=2$ and
$N_H=6\times10^{20}$ cm$^{-2}$ are assumed.}
\tablenotetext{b}{Most probable value, for details see Sec. \ref{sec:data}.}
\tablenotetext{c}{The nucleus; because of pileup, counts and flux are not true 
values.}
\end{deluxetable}

\begin{deluxetable}{crcrrrrr}
\tablecaption{\chandra~sources with optical counterparts.\label{tab:counter}}
               
\tablewidth{0pt}
\tablehead{
\colhead{Source} & \colhead{Unc.} & \colhead{Counterpart} & \multicolumn{3}{c}{Magnitude/Colors}
& \colhead{Remarks} & \colhead{Reference}\\ 
\colhead{CXO J} & \colhead{"} & \colhead{RA Dec} & \colhead{V} & \colhead{B-V} & \colhead{U-B} & \colhead{} &\colhead{}
}
\startdata
013253.4+303817&  6.1 & 01:32:53.30 +30:38:21.0 & 19.33 & -0.14 & -0.59 &         & [IFM93]\\

013253.9+303312&  8.0 & 01:32:53.62 +30:33:06.4 & 19.62 &  1.51 &  --   &M33d-676 & [M98]\\

013256.3+303558& 15.4 & 01:32:56.10 +30:35:52.0 & 18.40 & -0.10 & -1.10 & B30     & [IFM93]\\
               &      & 01:32:56.50 +30:35:50.0 & 19.27 & -0.31 & -0.30 &         & [IFM93]\\
               &      & 01:32:56.33 +30:35:52.4 & 18.19 & -0.09 & -0.89 & B1Ia/B30& [MAP95]\\
               &      & 01:32:56.71 +30:35:49.5 & 19.44 & -0.16 & -1.02 & O8-9    & [MAP95]\\
               &      & 01:32:57.13 +30:36:05.1 & 20.24 & -0.13 & -1.11 &         & [MAP95]\\
               &      & 01:32:56.41 +30:35:49.5 & 19.26 &  1.80 & --    &M33d-1785& [M98]\\
\enddata
\tablecomments{[IFM93] \citet{ivanov:93}, [MAP95] \citet{massey:95}, [MBH96]
\citet{massey:96}, [RW93] \citet{regan:93}, [M98] \citet{massey:98a}}
\end{deluxetable}

\begin{deluxetable}{crrrrr}
\tablecaption{\chandra-ROSAT matches. ROSAT data from \citep{haberl:01}.\label{tab:rosat}}
\tablewidth{0pt}
\tablehead{
\multicolumn{2}{c}{\chandra~source} & \multicolumn{4}{c}{ROSAT source}\\
\colhead{} & \colhead{Uncert.} & \colhead{Coordinates} & \colhead{Uncert.} & \colhead{Name} & \colhead{HRI rate}\\
\colhead{CXO J} & \colhead{arcsec} & \colhead{RA Dec} & \colhead{arcsec} &\colhead{} & \colhead{[$10^{-4}$cts/s]}
}
\startdata
013253.4+303817 &  6.1 & 01:32:53.400 +30:38:15.00 & 2.3 &  R027 &  20.10 $\pm$	 1.90\\
013253.9+303312 &  8.0 & 01:32:53.800 +30:33:10.00 & 2.6 &  R028 &   9.56 $\pm$	 1.30\\
013308.3+304802 &  4.8 & 01:33:08.700 +30:47:52.00 & 9.8 &  R040 &                   \\
013315.1+305317 &  8.2 & 01:33:15.100 +30:53:23.00 & 2.4 &  R049 & 188.00 $\pm$	12.00\\
013323.8+303517 &  3.2 & 01:33:23.800 +30:35:16.00 & 4.2 &  R054 &   1.56 $\pm$	 0.43\\
013324.4+304403 &  1.5 & 01:33:24.300 +30:44:04.00 & 2.0 &  R055 &  67.00 $\pm$	 1.80\\
013327.7+304647 &  3.8 & 01:33:27.500 +30:46:51.00 & 5.4 &  R057 &   2.01 $\pm$	 0.54\\
013327.9+303135 &  6.0 & 01:33:27.800 +30:31:30.00 & 6.0 &  R058 &   2.22 $\pm$	 0.60\\
\enddata

\end{deluxetable}

\begin{deluxetable}{crrrr}
\tablecaption{\chandra-XMM matches. XMM-Newton data from \citep{pietsch:04}.\label{tab:xmm}}
\tablewidth{0pt}
\tablehead{
\multicolumn{2}{c}{\chandra~source} & \multicolumn{3}{c}{XMM source}\\
\colhead{} & \colhead{Uncert.} & \colhead{Coordinates} & \colhead{Flux} & \colhead{X-ray ID}\\
\colhead{CXO J} & \colhead{arcsec} & \colhead{RA Dec} &  \colhead{[erg cm$^{-2}$ s$^{-1}$]} &\colhead{}
}
\startdata
013253.4+303817 &  2.6 & 1:32:53.46 30:38:14.9 & 1.32e-13 $\pm$	2.9e-15  &  X-1,SB1,LCB5,HP27  \\
013253.9+303312 &  3.0 & 1:32:53.82 30:33:12.1 & 1.14e-13 $\pm$	2.8e-15  &  X-2,SB2,LCB6,HP28  \\
013308.3+304802 &  2.5 & 1:33:08.21 30:48:04.3 & 1.12e-14 $\pm$	1.1e-15  &  LCB10,HP40  \\
013315.1+305317 &  0.7 & 1:33:15.03 30:53:17.9 & 1.06e-12 $\pm$	1.3e-14  &  X-4,SB5,LCB13,HP49  \\
013321.7+303858 &  1.4 & 1:33:21.66 30:38:58.8 & 3.85e-15 $\pm$	7.3e-16  &    \\
013321.9+303921 &  1.4 & 1:33:22.02 30:39:22.3 & 3.05e-15 $\pm$	7.1e-16  &    \\
013323.8+303517 &  1.6 & 1:33:23.86 30:35:18.0 & 6.66e-15 $\pm$	7.7e-16  &  HP54  \\
013323.9+304821 &  3.2 & 1:33:23.79 30:48:20.1 & 1.03e-14 $\pm$	1.1e-15  &    \\
\enddata
\end{deluxetable}

\normalsize

\end{document}